\RequirePackage[2020-02-02]{latexrelease}
\documentclass[aps, prd, twocolumn, showpacs, a4paper,
nofootinbib, superscriptaddress,longbibliography, floatfix]{revtex4-2}
\usepackage{blindtext}
\usepackage{graphicx}
\usepackage{epsfig}
\usepackage{bm}
\usepackage{amsfonts}
\usepackage{pgf}
\usepackage{color}
\usepackage[T1]{fontenc}
\usepackage[latin1]{inputenc}
\usepackage[english]{babel}
\usepackage{amsmath}
\usepackage{amssymb}
\usepackage{cases}
\usepackage{makecell}
\usepackage{hyperref}
\usepackage[draft,deletedmarkup=xout]{changes}
\usepackage{mathtools}
\usepackage{xcolor}
\usepackage{multirow}
\usepackage[version=4]{mhchem}
\usepackage{soul}
\definecolor{green}{rgb}{0.19,0.64,0.54}
\definecolor{blue}{rgb}{0,0,1}
\definecolor{reddish}{rgb}{0.65, 0.2, 0.2}
\definecolor{darkgreen}{rgb}{0.2,0.7,0.3}
\definecolor{darkblue}{rgb}{0.3,0.40,0.48}
\definecolor{gray}{rgb}{.8,.8,.8}

\hypersetup{,
	pdftitle={CVOS},
  	pdfauthor={I. Rybak et al.},
    colorlinks=true,       
    linkcolor=darkblue,          
    citecolor=darkgreen,        
    filecolor=reddish,      
    urlcolor=blue,           
    linktocpage=true
}
\usepackage{wrapfig}
\usepackage{ulem}

\graphicspath{ {./Figures/} }

\newcommand{\dd}{\mathrm{d}}

\begin{document}

\title{Cosmological evolution of Witten superconducting string networks}

\author{I. Yu. Rybak}
\email[]{Ivan.Rybak@astro.up.pt}
\affiliation{Centro de Astrof\'{\i}sica da Universidade do Porto, Rua
  das Estrelas, 4150-762 Porto, Portugal}
\affiliation{Instituto de Astrof\'{\i}sica e Ci\^encias do Espa\c co,
  CAUP, Rua das Estrelas, 4150-762 Porto, Portugal}

\author{C. J. A. P. Martins}
\email{Carlos.Martins@astro.up.pt}
\affiliation{Centro de Astrof\'{\i}sica da Universidade do Porto, Rua
  das Estrelas, 4150-762 Porto, Portugal}
\affiliation{Instituto de Astrof\'{\i}sica e Ci\^encias do Espa\c co,
  CAUP, Rua das Estrelas, 4150-762 Porto, Portugal}

\author{Patrick Peter}
\email{peter@iap.fr}
\affiliation{${\cal G}\mathbb{R}\varepsilon\mathbb{C}{\cal O}$ --
  Institut d'Astrophysique de Paris, CNRS \& Sorbonne Universit\'e,
  UMR 7095 98 bis boulevard Arago, 75014 Paris, France}
\affiliation{Centre for Theoretical Cosmology, Department of Applied
  Mathematics and Theoretical Physics, University of Cambridge,
  Wilberforce Road, Cambridge CB3 0WA, United Kingdom}

\author{E. P. S. Shellard}
\email{E.P.S.Shellard@damtp.cam.ac.uk}
\affiliation{Centre for Theoretical Cosmology, Department of Applied
  Mathematics and Theoretical Physics, University of Cambridge,
  Wilberforce Road, Cambridge CB3 0WA, United Kingdom}

\begin{abstract}
We consider the evolution of current-carrying cosmic string networks
described by the charge-velocity-dependent one scale (CVOS) model
beyond the linear equation of state regime, specifically focusing on
the Witten superconducting model. We find that, generically, for
almost chiral currents, the network evolution reduces dynamically to that of the
linear case, which has been discussed in our previous work. However,
the Witten model introduces a maximum critical current which
constrains the network scaling behaviour during the radiation era when
currents can grow and approach this limit.  Unlike the linear model,
only if the energy density in the critical current is comparable to
the bare string tension will there be substantial backreaction on the
network evolution, thus changing the observational predictions of
superconducting strings from those expected from a Nambu-Goto network.
During the matter era, if there are no external sources, then
dynamical effects dilute these network currents and they disappear at
late times.
\end{abstract}

\date{\today}
\maketitle

\section{Introduction}

In Ref.~\cite{Kibble}, Tom Kibble proposed that one-dimensional
topological defects, dubbed cosmic strings, should form in many
extensions of the standard model of particle physics. This has been
confirmed explicitly in many phenomenological studies~(see, for
example, refs~\cite{JeannerotRocherSakellariadou, Allys,SarangiTye}).
A detailed examination of most of these scenarios indicates that the
strings should also be endowed with superconducting currents, as
proposed originally by Witten~\cite{Witten:1984eb}, which have many
interesting consequences ~(e.g., refs~\cite{Davis:1988ij,
  Peter:1992ta, Davis:1995kk, Bin_truy_2004, Allys2,
  AbeHamadaYoshioka, FukudaManoharMurayamaTelem,
  PhysRevD.107.L031903}). Beyond the original bosonic superconducting
model, a variety of different mechanisms have also been proposed to
generate string
currents~\cite{Witten:1984eb,Everett,KibbleLozanoYates,Lilley:2010av},
making superconductivity even more ubiquitous.

The presence of a superconducting current flowing along the strings
necessarily influences the corresponding cosmic string network
evolution.  Recently, we developed a charge-velocity-dependent
one-scale (CVOS) model~\cite{Martins:2020jbq}, which extends previous
work~\cite{Martins:1996jp, Martins:2000cs, Oliveira:2012nj,
  Vieira:2016, Rybak:2017yfu} and offers an analytical approach to
describing the most relevant statistical features of the network
evolution. This approach includes phenomenological parameters that
could, in the non-current-carrying case, be directly measured or
statistically inferred from numerical simulations, thereby calibrating
the models. In principle, such calibrated models can be used to
evaluate the stochastic background of gravitational waves produced by
a current-carrying network, much in the same way as is done for bare
string networks~\cite{Auclair:2022ylu, RybakSousa2}, as well as other
observational signatures.  However, before one can make reliable
predictions, there is a further bottleneck: network simulations for
superconducting strings are not yet available with which to reliably
estimate the CVOS model parameters, so in this case we need to survey
a wider parameter range of possible physical consequences.

The CVOS analytic model requires a specific equation of state
appropriate for each particular field theory model for
current-carrying cosmic strings. Having reviewed first the general
equation of state~\cite{Martins:2020jbq} for cosmic string network
evolution, and then specialised to the simplest linear
case~\cite{Martins:2021cid}, we now focus, in the present paper, on
the original Witten model ~\cite{Witten:1984eb}---or more precisely
the neutral Witten model, containing no long-range
electromagnetic-like interactions~\cite{Peter:1992dw,
  Peter:1992ta}.

This paper is organised as follows: in Sec.~\ref{GenCVOS}, the
CVOS model equations are reviewed, with a particular emphasis put on
the various charge and current loss mechanisms, either in the
process of loop formation or due to local curvature; we close
the system by assuming that
the framework is that of radiation or matter dominated Friedman-Lema\^{i}tre-Robertson-Walker (FLRW) background. The Witten model
equation of state is implemented in Sec.~\ref{WittenEOS}. It
entails the existence of a critical current, leading to a 
leakage parameter dependence in the total charge.

The core of the article is
Sec.\ref{Evolution}, in which we study specific
network dynamics during cosmological evolution. We demonstrate that for
small currents the evolution can be described by the previous linear
model. However, for regimes with growing currents in the radiation era
we note the important role of the critical current in determining
scaling solutions. We end with some conclusions
and a discussion of the outcomes, their
susceptibility to our underlying assumptions and potential
observational implications.

\section{Generalised CVOS model}
\label{GenCVOS}

In what follows, we make use of the recently developed CVOS model to
describe the most relevant statistical properties of a network of
superconducting cosmic strings, within the thin-string approximation;
we refer the reader to Ref.~\cite{Martins:2020jbq} in which the
relevant calculations are detailed.

\subsection{Relevant thermodynamical variables}

Upon averaging over all the long strings, the current-carrying string
network ends up being characterized by four macroscopic variables,
namely the root-mean-square (RMS) velocity $v$ of the strings, the
energy density $\rho$, the charge $Y$ and the 4-current amplitude $K$
(sometimes also called chirality, as it measures the spacelike or
timelike character of the integrated current). These variables are
originally defined in terms of the root mean squares of the timelike
and spacelike currents, respectively $Q^2 = \langle q^2 \rangle$ and
$J^2 = \langle j^2 \rangle$, namely
\begin{equation}
Y = \frac12\left( Q^2 + J^2 \right) \ \ \hbox{and} \ \ 
K =  Q^2 - J^2,
\label{Q2K2}
\end{equation}
so that $Q^2 = Y+K/2$ and $J^2 = Y-K/2$.
Note that $Y$ being positive definite, Eq.~\eqref{Q2K2}
implies that the constraint
\begin{equation}
|K| \leq 2 Y
\label{K2Y}
\end{equation}
should be satisfied at all times.

The Lorentz-invariant microscopic chirality $\kappa = q^2-j^2$ is what
enters the surface Lagrangian $f(\kappa)$ from which the equations of
motion are derived. Averaging this quantity suggests its replacement
with a macroscopic version, $F(K) = \langle f(\kappa)\rangle$, which,
for simplicity, we assume to take the same form, as we discuss below.

The energy density $\rho$ can be split into two contributions, namely
that coming from the bare (without current) strings $\rho_0$, and that
coming from the current itself. The Brownian assumption allows us to
rewrite this energy density through two conformal characteristic
lengths $L_\textsc{c}$ and $\xi_\textsc{c}$, leading to
\begin{equation}
    \label{Brownian}
    \rho = \frac{\mu_0}{L_\textsc{c}^2 a^2} \quad 
    \text{and} \quad \rho_0
    = \frac{\mu_0}{\xi_\textsc{c}^2 a^2},
\end{equation}
where $\mu_0$ is the bare string tension, $L_\textsc{c}$ and
$\xi_\textsc{c}$ correspond to the total (current-carrying) and bare
string networks respectively, and $a$ is the scale factor of the FLRW
cosmological solution
\begin{equation}
\dd s^2 = a^2(\tau) \left( \dd \tau - \dd \bm{x}^2 \right),
\label{FLRW}
\end{equation}
$\tau$ being the conformal time. The relation between the
characteristic lengths is found to be
\begin{equation}
\xi_\textsc{c} = \sqrt{F-2Q^2 F'} L_\textsc{c} = W L_\textsc{c},
\label{Lxi}
\end{equation}
thereby defining $W$ (with $F'\equiv \dd F/\dd K$). For later
reference, we note that
 \begin{equation}
\frac{\dot{W}}{W} = -\frac{1}{W^2} \left[ \dot{Y} F'+ \left( 
Y+\frac{K}{2}\right)\dot{K} F'' \right],
 \label{dotWW}
 \end{equation}
with the overdot henceforth denoting conformal time derivatives.

\subsection{Charge and current leakage}

There are two mechanisms by which the superconducting string network
can lose energy: loop production and charge leakage.

One can phenomenologically describe charge and current losses through
leakage by demanding that the larger the charge or current, the larger
the loss (we generalise here the description proposed in
Ref.~\cite{Martins:2021cid}).  This translates into
\begin{equation}
    \left. \frac{\dd Q^2}{\dd \tau} \right|_\text{leak} =
    - A \frac{Q^2}{\xi_\textsc{c}} \quad \text{and} \quad
    \left. \frac{\dd J^2}{\dd \tau} \right|_\text{leak} =
    - B \frac{J^2}{\xi_\textsc{c}},
\label{chargeLeak}
\end{equation}
where $A$ and $B$ are called respectively the charge and current
leakage efficiencies and the subscript "leak" indicates we restrict
attention to the part specifically due to leakage. Setting $A_{_\pm} =
A\pm B$, Eq.~\eqref{chargeLeak} implies the time evolution
contribution for the variables $Y$ and $K$,
\begin{equation}
\begin{gathered}
\left. \frac{\dd Y}{\dd \tau} \right|_{\text{leak}} = 
-\frac{1}{2\xi_\textsc{c}} \left( A_{_+} Y + A_{_-}
\frac{K}{2}\right), \\
\left. \frac{\dd K}{\dd \tau} \right|_{\text{leak}} = 
-\frac{1}{\xi_\textsc{c}} \left( A_{_-} Y + A_{_+} \frac{K}{2}\right).
\end{gathered}
\label{dYKleak}
\end{equation}
The behaviour of the phenomenological parameters $A_\pm$ are not known
precisely, so we cannot \textit{a priori} set them to constants, though we
have made this linear assumption previously \cite{Martins:2021cid}. We shall see below for the
Witten model that, because the physically relevant equation of state
entails a critical current, leakage should increase as the current
approaches criticality.  In any case, the specific form of the charge
leakage efficiencies has to be inferred from field theoretic studies
of microscopic string behaviour and future high-resolution numerical
simulations of evolving random networks.

Since the non-current-carrying contribution of the network energy $\rho_0$ is
of course insensitive to these losses, i.e.,
\begin{equation}
\left. \frac{\dd \rho_0}{\dd \tau} \right|_{\text{leak}} = 0,
\label{rho0leak}
\end{equation}
it follows that Eq.~\eqref{rho0leak}, together with the definition
\eqref{Brownian}, implies that $\dot{\xi}_\text{c,leak} =0$, which,
upon using \eqref{Lxi}, yields
\begin{equation}
\left. \frac{\dot{L}_\textsc{c}}{L_\textsc{c}} \right|_\text{leak}
= -\left. \frac{\dot{W}}{W} \right|_\text{leak}.
\label{Lcdotleak}
\end{equation}
Since $F$ depends only on the chirality, one has $\dot{F} = \dot{K}
F'$, and one gets the restriction of Eq.~\eqref{dotWW} due to leakage
by substituting \eqref{dYKleak} into \eqref{dotWW},
which provides the correction to the equation of motion of the
characteristic length $L_\textsc{c}$.

\subsection{Loop chopping}

The energy loss due to the production of loops from long strings takes
the form~\cite{KIBBLEEvolut}
\begin{equation}
    \label{chopping}
    \left. \frac{\dd \rho_0}{\dd \tau} \right|_{\mathrm{loops}}
    =  -\frac{c v \rho_0}{\xi_\textsc{c}}, \quad
    \left. \frac{\dd \rho}{\dd \tau}  \right|_{\mathrm{loops}} =
    - g(Q,J) \frac{c v \rho}{\xi_\textsc{c}},
\end{equation}
where the subscript "loops" means we restrict attention, in the
calculation of the time derivative, to the contribution due to loop
production on the long string densities.

In Eq.~\eqref{chopping}, $c$ is the loop chopping efficiency, which will be
set to its Nambu-Goto value $c_\textsc{ng} \sim 0.23$ in the
forthcoming numerical calculation, and $g(Q,J)$ represents the
modification of this bare chopping efficiency to account for the
effects of the charge. Lacking numerical simulations for
current-carrying strings, one can first assume the loop production to
not be significantly modified by inclusion of current effects; this
assumption will have to be tested when current-carrying string network
simulations become available.

Using the Brownian network properties~\eqref{Brownian} with the loop production equations~\eqref{chopping} implies
\begin{equation}
    \left. \frac{\dot{\xi}_\textsc{c}}{\xi_\textsc{c}}
    \right|_{\mathrm{loops}} = \frac{cv}{2\xi_\textsc{c}} \quad
    \mathrm{and} \quad \left. \frac{\dot{L}_\textsc{c}}{L_\textsc{c}}
    \right|_{\mathrm{loops}} = g\frac{cv}{2\xi_\textsc{c}}\,.
\label{choppingnew}
\end{equation}
Since \eqref{Lxi} yields
$$
\frac{\dot{\xi}_\textsc{c}}{\xi_\textsc{c}} =
\frac{\dot{L}_\textsc{c}}{L_\textsc{c}} +\frac{\dot{W}}{W},
$$
we obtain the relation 
\begin{equation}
    \left. \frac{\dot{W}}{W} \right|_{\mathrm{loops}}
    =-\frac{cv}{2\xi_\textsc{c}} (g-1),
\label{chopW}
\end{equation}
which means that if $g \neq 1$, the loop production affects the $Q^2$
and $J^2$ parameters.  Here, we again assume that the loop production function $g$
has a linear dependence on $Q^2$ and $J^2$, which implies that
\begin{equation}
\begin{gathered}
\label{chopQJ}
    \left. \frac{ \dd Q^2 }{ \dd \tau} \right|_{\mathrm{loops}}
    =-g_{_Q}\frac{c v}{\xi_\textsc{c}} Q^2 , \\
    \left. \frac{ \dd J^2 }{ \dd \tau} \right|_{\mathrm{loops}}
    =-g_{_J}\frac{c v}{\xi_\textsc{c}} J^2 ,
\end{gathered}
\end{equation}
where $g_{_Q}$ and $g_{_J}$ are some constants that tell us how much
time-like and space-like components of currents are lost due to loops
production. One can demonstrate that these constants are related to
the function $g$ in Eq.~\eqref{choppingnew} by the following expressions
\begin{equation}
    \label{ggQgJ}
    g = 1 - g_{_Q} \frac{F^{\prime}+2 Q^2 F^{\prime \prime} }{ F-2 Q^2
      F^{\prime}} Q^2 - g_{_J} \frac{F^{\prime}-2 Q^2 F^{\prime
        \prime}}{ F-2 Q^2 F^{\prime}} J^2.
    \end{equation}
We note that this improved parametrization of the function $g$ offers
some further clarity over that used in
previous approaches \cite{Oliveira:2012nj, Rybak:2017yfu,
  Martins:2020jbq, Martins:2021cid}. Due to limited knowledge
about the form of $g$ in the equation for $L_\textsc{c}$, we
assume that it should be linearly proportional to the time-like and
space-like components in the corresponding current loss functions.

Collecting time-like and space-like components, one obtains
expressions for the charge $Y$ and 4-current amplitude $K$ in the
following forms
\begin{equation}
    \begin{gathered}
        \label{LoopsYK}
    \left. \dot{Y} \right|_{\mathrm{loops}} = - \frac{c
      v}{2\xi_\textsc{c}} \left( g_{_+} Y +
    g_{_-}\displaystyle\frac{K}{2} \right) , \\ \left. \dot{K}
    \right|_{\mathrm{loops}} = - \frac{c v}{\xi_\textsc{c}}
    \left(g_{_-} Y + g_{_+} \displaystyle\frac{K}{2} \right), \\
    \end{gathered}
\end{equation}
where $g_{_\pm} = g_{_Q} \pm g_{_J}$.

\begin{widetext}

\subsection{Equations of motion}

The relevant equations of motion for the quantities of cosmological
interest, namely the characteristic length of the network
$\xi_\text{c}$, the root mean square velocity $v$, together with the
charge $Y$ and chirality $K$ can now be written after collecting all
the modifications which must be added to Eqs.~(43) of
Ref.~\cite{Martins:2020jbq} and discussed above. First, we set the
characteristic length $L_\textsc{c}$ to behave as $L_\textsc{c} =
\zeta(\tau) \tau$, so that a scaling solution will correspond to a
constant value of the fraction $\zeta$. Taking into account the
mechanisms described above, one finds
\begin{equation}
 \dot{\zeta} \tau = \frac{ n \zeta }{W^2} \left[ v^2 \left(F -
   F^{\prime} K\right) - 2 Y F^{\prime} \right] + \frac{c v g}{2 W} -
 \frac{F^{\prime}}{2 W^3} \left( A_{_+} Y + A_{_-} \frac{K}{2} \right)
 - \frac{F^{\prime \prime}}{W^3} \left( Y + \frac{K}{2} \right) \left(
 A_{_-} Y + A_{_+} \frac{K}{2} \right) - \zeta ,
\label{zetadot}
\end{equation}
where $g$ is given by the Eq.~\eqref{ggQgJ}. It turns out however
that, using the relation \eqref{Lxi} to express this current-carrying
network correlation length in terms of the bare one $\xi_\textsc{c}$,
and substituting into \eqref{zetadot} all the results obtained in the
previous section, including the time development of the charge $Y$ and
chirality $K$, one finds that the bare correlation length provides a
tremendous simplification to the equations describing the system
evolution. Indeed, setting $\xi_\textsc{c} = \epsilon \tau$, this
system becomes
\begin{subequations}
\label{CVOS}
\begin{align}
\label{CVOSzeta_correction}
\dot{\epsilon} \tau = &\ \frac{1}{W^2} \left[ n \epsilon v^2 \left( F
  - F^{\prime} K \right) - 2 v k Y F^{\prime}\right] + \frac12 c v
-\epsilon, \\
\label{CVOSv} \dot{v} \tau  = &\ \frac{1-v^2}{W^2} \left\{
\frac{k}{\epsilon} \left[ F + 2 \left( Y - \frac{K}{2} \right)
  F^{\prime} \right] - 2 v n \left( F - F^{\prime} K \right) \right\},
\\
\label{CVOSY} \dot{Y} \tau = &\ \left(
\frac{v k}{\epsilon} -n \right) \frac{ 2 Y F^{\prime}+(4
  Y^2-K^2)F^{\prime \prime}}{F^{\prime}+ (2 Y+K) F^{\prime \prime}} -
\frac{c v}{2 \epsilon} \left( g_{_+} Y + g_{_-} \frac{K}{2} \right) -
\frac{2 A_{_+} Y + A_{_-} K}{4 \epsilon}, \\
\label{CVOSK} \dot{K} \tau = &\ 2  \left(
\frac{v k}{\epsilon} - n \right) \frac{F^{\prime} K}{F^{\prime}+(2
  Y+K) F^{\prime \prime}} - \frac{c v}{\epsilon} \left( g_{_-} Y +
g_{_+} \frac{K}{2} \right) - \frac{2 A_{_-} Y + A_{_+} K}{2 \epsilon}.
\end{align}
\end{subequations}
\end{widetext}
Eq.~\eqref{CVOSzeta_correction} is indeed much simpler than
\eqref{zetadot}, as announced. We have checked that one could solve
either the system \eqref{CVOS} and substitute \eqref{Lxi} into the
solution, or only the last three, i.e. \eqref{CVOSv}, \eqref{CVOSY}
and \eqref{CVOSK} in terms of $L_\textsc{c}$ together with
\eqref{zetadot}. Both solutions are numerically identical, as they
should.

A scaling solution is then achieved whenever the functions of time
$\zeta(\tau)$, $v(\tau)$, $Y(\tau)$ and $K(\tau)$ simultaneously evolve
towards constant solutions $\zeta_\textsc{sc}$, $v_\textsc{sc}$, 
$Y_\textsc{sc}$ and $K_\textsc{sc}$: the characteristic length
is then a constant fraction of the Hubble scale.

In Eqs.~\eqref{CVOS}, the so-called momentum parameter $k$ is assumed
to be a function of the RMS velocity $v$ only, and is defined
through~\cite{Martins:2000cs}
\begin{equation}
    \label{MomentK}
    k \equiv k(v) = \frac{2 \sqrt{2}}{\pi} \frac{1-8 v^6}{1+8 v^6}
    \left( 1-v^2 \right) \left( 1 + 2 \sqrt{2} v^3 \right),
\end{equation}
which is the same as was found to describe well the Nambu-Goto network
simulations. In other words, we assume that the explicit form of this
function, which comes mostly from curvature effects along the string
worldsheets, is not affected by the presence of non-zero charge or
chirality.  Note that there is however an implicit dependence, because
their presence does impact the RMS velocity.

\section{Witten Equation of state}
\label{WittenEOS}

To specify a particular type of superconducting cosmic string, we need
to define its equation of state. The Witten model, originally proposed
in Ref.~\cite{Witten:1984eb}, or rather its neutral version
\cite{Peter:1992dw} (to ensure long-range electromagnetic-like effects
to be negligible \cite{Peter:1992ta}) was found to be accurately
characterised by the following averaged equation of
state~\cite{Carter:1994hn}

\begin{subnumcases}{\label{RealModel}}
F_{\text{mag}}(K) = 1 - \displaystyle\frac{1}{2} \frac{K}{1 - \alpha
  K} & \text{for } $K \leq 0$, \label{Fmag}\\
F_{\text{elec}}(K) = 1 + \displaystyle\frac{\ln (1- 2\alpha K) }{4
  \alpha} & \text{for } \; $K \geq 0$, \label{Felec}
\end{subnumcases}
where the model-dependent dimensionless parameter $\alpha$ is given by
$\alpha = (m_\textsc{h}/m_\sigma)^2$, with $m_\sigma$ the vacuum mass
of the current-generating condensate and $m_\textsc{h}$ that of the
string-forming Higgs field. Generically, $\alpha$ is expected to be
much larger than unity, and some degree of fine-tuning would be
required to obtain ${\cal{O}}(1)$. In any case, it must lie in the
range $1<\alpha<\infty$, to which we will restrict our attention
below. Note that the coefficient in \eqref{Felec} differs in an
irrelevant way from that in \cite{Martins:2020jbq}: this is to ensure
the small $K$ behavior to be identical for both functions up to second
order.

\begin{figure}[t]
\begin{center}
\includegraphics[scale=0.3]{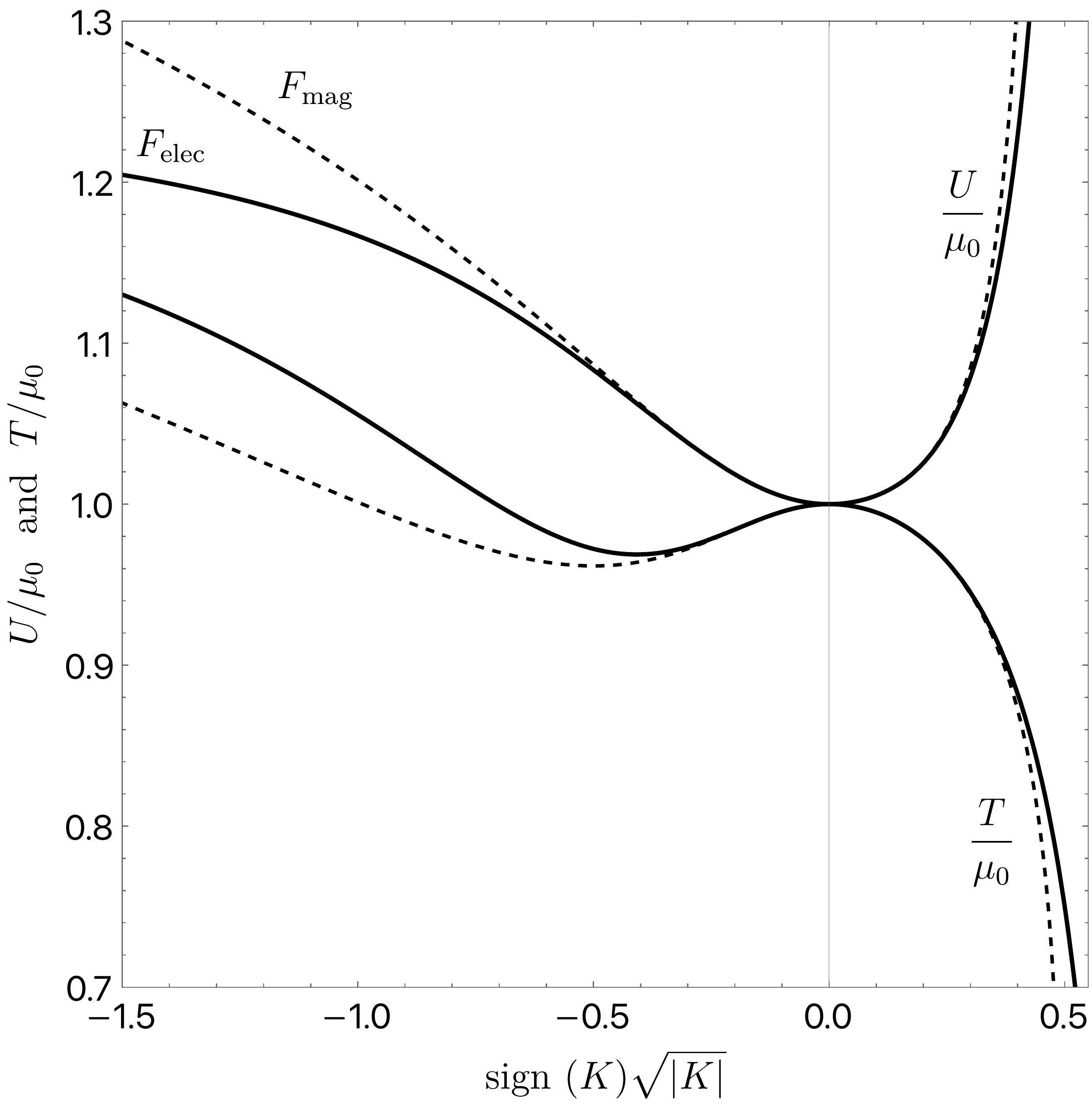}
\caption{Equation of state for the Witten model using
  Eq.~\eqref{RealModel} with $\alpha=2$. The constraints on the state
  parameter $K$ are clearly seen: on the magnetic side with $K\leq 0$,
  below the critical value $-1/(3 \alpha)$, the energy per unit length
  $U$ and the tension $T$ are both decreasing, implying the
  longitudinal perturbation velocity $c_\textsc{l}^2 = -\dd T/\dd U <
  0$, while on the electric regime with $K\geq 0$, the ever decreasing
  tension eventually reaches a point for which $T < 0$, and hence a
  transverse perturbation velocity $c_\textsc{t}^2 = T/U < 0$: both
  cases lead to instabilities and are thus dynamically
  excluded.\label{fig:eos} }
\end{center}
\end{figure}

Shown in Fig.~\ref{fig:eos} is a representation of the equation of
state proposed in Eq.~\eqref{RealModel}, which also implies the
following restriction on the 4-current amplitude value
\cite{Peter:1992dw}
\begin{equation}
\label{K_Restr}
-\frac{1}{3 \alpha} < K < \frac{1-\text{e}^{-4 \alpha}}{2 \alpha}.
\end{equation}
Here the first constraint stems from the requirement that the
longitudinal velocity of perturbations propagating along the
worldsheet be positive in the magnetic case with $K\leq 0$, i.e. using
$F_\text{mag}$ \eqref{Fmag}. The second constraint merely expresses
$F_\text{elec} \geq 0$ \eqref{Felec}. These limits lead to $K
\rightarrow 0$ as $\alpha \to \infty$, meaning that if the
current-carrier mass is vanishingly small compared to that of the
Higgs field, the contribution of the current becomes negligible and
one recovers a Nambu-Goto string. Note also that for small
chiralities, Eq.~\eqref{RealModel} can be expanded as $F \sim 1 -
\frac12 K - \frac12 \alpha K^2$ in both cases.

It is important to emphasise that the equation of
state~\eqref{RealModel} is applicable only provided the current
remains in the range given by Eq.~\eqref{K_Restr}. If for some reason
(due to initial conditions or dynamical evolution) the chirality came
to exceed this range, one should include possible electromagnetic
corrections \cite{GanguiPeterBoehm} that might change the equation of
state or, most probably, lead to additional charge loss mechanisms.
(We briefly discuss one such example in Sect. \ref{Beyond}.)  However,
in this work we restrict our attention to macroscopic variables and we
can anticipate that, on average, the equation of
state~(\ref{RealModel}) holds.  In that case, and for the relevant
$\alpha\gg 1$, the string network on average should tend towards
chiral conditions, i.e $K_{\rm sc} = 0$. The largest deviation from
the chiral case happens when $\alpha\simeq 1$, implying $-0.33
\lesssim K_{\rm sc} \lesssim 0.86$.  This $\alpha \sim 1$ regime is
actually the most interesting case because it is where the
backreaction on the underlying string is most significant to the
extent that it may even lead to stable vorton solutions.

The equation of state \eqref{RealModel}, which implies the existence
of a maximum chirality $K_{\rm cr} \sim \mathcal{O}(1/\alpha)$
from \eqref{K_Restr}, will also have a dramatic effect on the
evolution of the string currents.  In principle, a critical current in
$K$ may not seem to impose a limit on the total current $Y$ because of
the Lorentz invariance along a straight string, which allows $Y$ to be
boosted up to arbitrarily large values.  However, in the realistic
context of an expanding background, we work in a special cosmological
frame in which the overall Brownian string network is at rest (like
the CMB).  The actual RMS current $Y$ consists of random contributions
of correlation length which, for the present Witten model,
consists of both positive and negative chirality currents limited in
magnitude by the critical current $|K| \lesssim K_\text{cr}$.  For
this reason, we can expect the averaged total current to obey 
approximately the same
limit $Y \lesssim K_\text{cr}$ and, in fact, the cumulative stochastic
current will be considerably smaller and depend on the ratio of the
correlation length of the current to that of the string and other
factors.  This limit would probably have to be determined
quantitatively using numerical simulations, but we can be confident that
the average current will not exceed much 
the critical current $K_{\rm cr}$.

In order to implement this limit phenomenologically, we can consider
what would happen when $Y > Y_\text{cr}$, which we assume is of order
$Y_\text{cr} \approx 2 K_\text{cr}$ because of the constraint
\eqref{K2Y}, representing an average
current in which some random microscopic regions must have chiral
currents $K$ that exceed the critical value.  In that case we
anticipate an additional enhancement of the charge leakage, along the
lines of the discussion in Ref.~\cite{GanguiPeterBoehm}, and in
agreement with the results of instabilities observed in numerical
simulations in both the electric (charge loss) and magnetic (current
unwinding) regimes (see e.g.\ \cite{LagunaMatzner,Lemperiere1}).
One may use the electric regime bound in \eqref{K_Restr}, in our
numerical calculations below, to assume $Y_\text{cr} =
2/(3 \alpha)$ for the sake of definiteness.

Whenever $Y \gtrsim Y_\text{cr}$, one expects a rapid escape of
particles and energy
from a localised string region due to the current unwinding or charge
emission.  We can model this enhancement of current or charge
loss by a modification of the charge leakage parameter $A_\pm$ in
\eqref{dYKleak} to become $Y$-dependent as follows
\begin{equation}
    \label{NonLinLeakage}
 A_{\pm}(Y) = \frac{A_{\text{const}}}{1-\text{e}^{-(Y -
     Y_\text{cr})^2} }.
\end{equation}
The nonlinear leakage coefficient $A_{\text{const}}\ge 0$
will ensure that the average
current $Y$ is constrained by the critical current cut-off
\eqref{K_Restr} at all times.  As we will see, this eliminates some of
the large-current scaling regimes found previously for the linear
model, unless the critical current itself is sufficiently large
$\alpha \sim 1$.

\section{Network evolution}
\label{Evolution}

Having defined the thermodynamical equations of
motion and the relevant equation of state, we now are in a
position to analyse the various cases for the macroscopic
variables $L_\textsc{c}$, $v$, $Y$ and $K$. We first consider
in what follows, Sec.~\ref{NoLeak}, the situation for which
the charge leakage is negligible before taking it into account
in a phenomenological way in Sec.~\ref{Beyond}.

\subsection{Neglecting charge leakage}
\label{NoLeak}

We begin by investigating the case when the phenomenological parameter
describing charge leakage mechanisms is absent, i.e. we set
$A_{\pm} = 0$. A
realistic current-carrying string network should lose some energy into
radiation, but the actual amount is unknown given the lack of relevant
simulation information, so we have to consider the possibility of
negligible losses

\begin{figure}[h!]
\begin{center}
\includegraphics[scale=0.38]{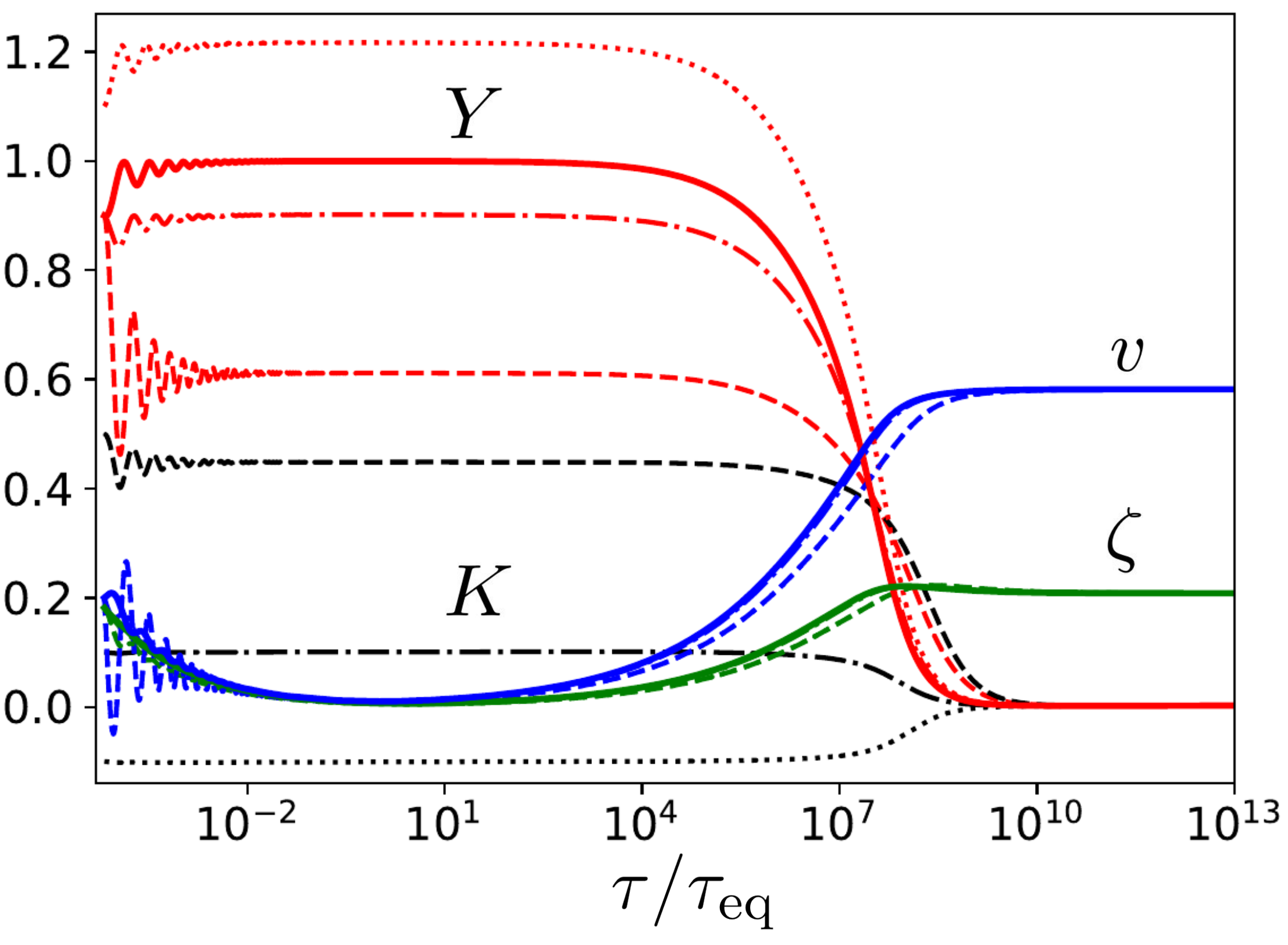}
\caption{\label{Figure:Frozen} Time evolution for the no-leakage case
  $A=0$ of the velocity $v$, charge $Y$, chirality $K$ and
  characteristic length ratio $\zeta$ as functions of conformal time,
  beginning deep in the radiation and ending in the matter era.  We
  chose the loop chopping parameter value to coincide with the Nambu-Goto
  case, i.e. $c=0.23$, and maximise the effect of the equation of
  state by setting $\alpha=1$. Shown here are the cases for initial
  conditions set with initial values $K_\text{ini}=0$ (full lines),
  $K_\text{ini}=0.5$ (dashed lines), $K_\text{ini}=0.1$ (dash-dotted
  lines) and $K_\text{ini}=-0.1$ (dotted lines).  For
  $K_\text{ini}=0$, and more generally for initial conditions close to
  the chiral case, one recovers the linear regime frozen solution with
  the saturation point (constant value of $Y$) depending on the
  scaling value of $K$ in a manner independent of the value of $\alpha
  >1$.}
\end{center}
\end{figure}

When the equation of state is of the linear kind, in practice when
$F\to 1-\frac12 K$, which is the small current (chiral) limit, it was
shown in Ref.~\cite{Martins:2021cid} that this leads to a `frozen'
network with $Y\to 1$, while the RMS velocity $v$ and the
correlation length ratio $\zeta$ decrease
as a power law with time during the radiation era.  As we
will see below, this behaviour does not persist in the full
Witten model unless there is a large critical current. In
the matter-dominated era, the network behaves in a Nambu-Goto way.
Fig.~\ref{Figure:Frozen} shows a typical numerical solution of
the set of Eqs.~\eqref{CVOS} with $A_{\pm}=0$ and using the Witten
equation of state \eqref{RealModel} showing the aforementioned time
developments.

When the initial conditions for the network, set deep into the
radiation era, are close to chirality, i.e. setting $K_\text{ini} \ll
1$, as is expected from the phase transition and the Kibble
current-forming mechanism, we found that there is no visible
difference between this model and the linear case, and this is true
for any value of $\alpha>1$.  If the initial value of the chirality
$K_\text{ini}$ is not negligible and positive (respectively negative),
the saturating value of the charge is $Y_\text{max} < 1$ (resp.
$Y_\text{max}> 1$); it is exemplified in Fig.~\ref{Figure:Frozen}.

In Ref.~\cite{Martins:2021cid}, it was found that the charge
saturation regime was leading to $Y_\text{max} \to 1$, a condition
entirely depending on the linearity of the equation of
state. Deviations of the charge saturation value $Y$ from unity for
the non-linear equation of state discussed in the present work
is due to the explicit appearance of $K$
in the relation \eqref{Lxi} between bare string energy and energy of
the current \eqref{Brownian}.  This effect increases with $\alpha$, so
in the limit of large mass difference, the current $K\to 0$, and
the saturation current $Y_\text{max} \sim 1$.  After the
radiation to matter transition, all solutions tend to the currentless
Nambu-Goto case.

For negligible charge losses, we conclude that the network behaviour
is not substantially modified by using the Witten equation of state
instead of the linear one. Nevertheless, there is one caveat to be
pointed out.  For large $\alpha$, even if the saturation current is
small, with $Y\ll 1$, one could still be in a regime with $\alpha Y >
1$ which would seem to exceed expectations on the existence of a
critical current for the string. Clearly, having $\alpha \gg 1$ and
$Y\sim 1$ seems unphysical (i.e. inconsistent with the realistic
equation of state we are studying). We revisit this point in
Sect. \ref{Beyond}.

\subsection{Charge leakage with critical current}
\label{Beyond}

Having discussed the modification due to the Witten equation
of state on the frozen network, we introduce the leakage and
its charge dependence \eqref{NonLinLeakage}. We first consider
the chopping efficiency to be independent of the charge and
current, so that $g\to 1$, i.e. assuming negligible biases
$g_{_Q}\ll 1$ and $g_{_J}\ll 1$, and then include their
contribution.

\subsubsection{The unbiased case}
\label{NoBias}

\begin{figure}[t]
\begin{center}
\includegraphics[scale=0.38]{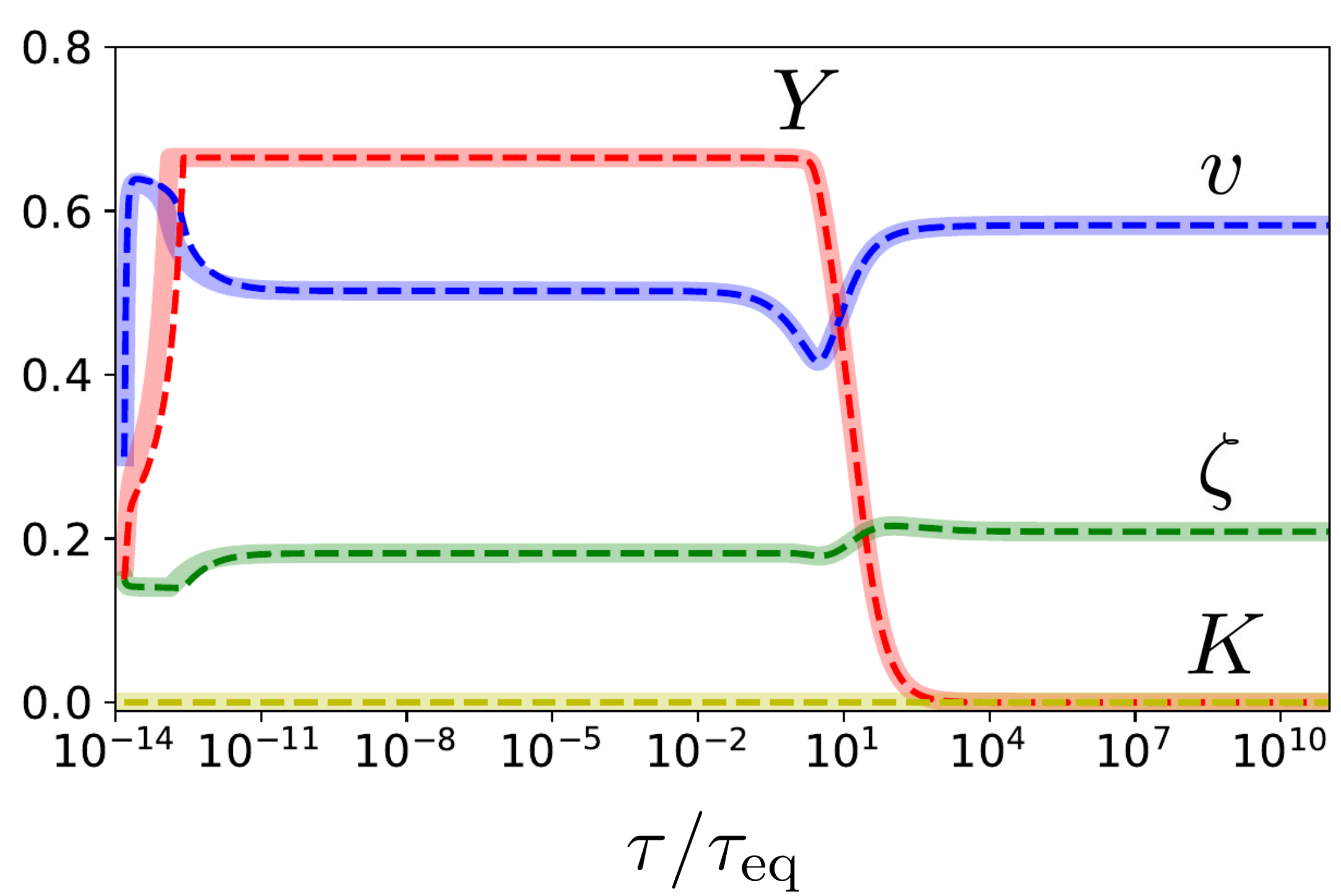}
\includegraphics[scale=0.38]{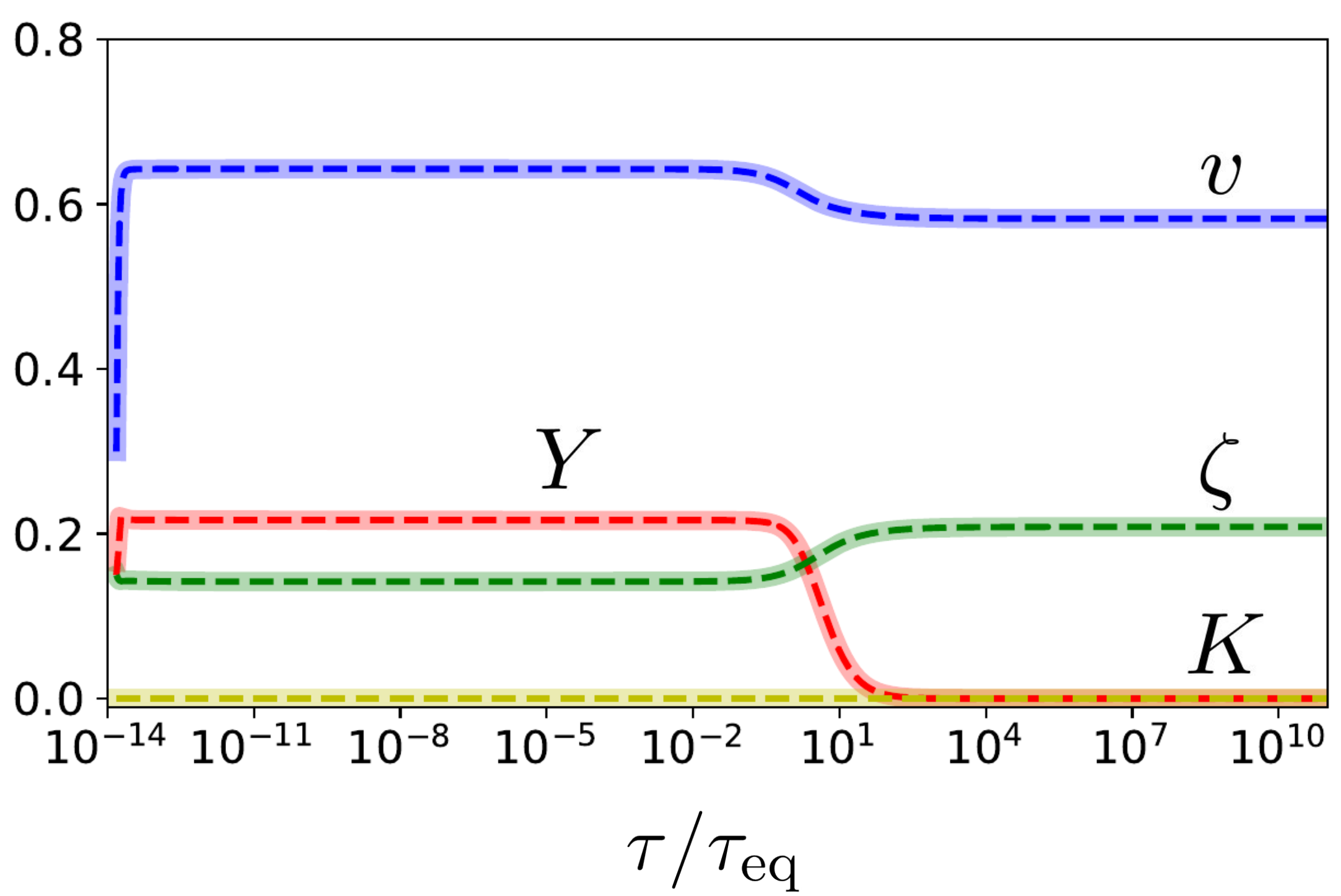}
\caption{\label{Figure:NonLinA_expN} Time evolution of the velocity
  $v$, charge $Y$, chirality $K$ and $\zeta$ through the radiation and
  matter epochs, using parameter values $c=0.23$, $g_{\pm}=0$,
  $A_{_-}=0$ and $A_{_+}$ given by \eqref{NonLinLeakage} or
  \eqref{NonLinLeakageInvers} with
  $A_{\text{const}}=10^{-6}$. Evolving using either the Witten
  \eqref{RealModel} (dashed lines) or the linear equation of state
  (solid line) for $\alpha=1$ (upper panel) and
  $\alpha=3$ (lower panel) lead to indistinguishable
  evolutions in this case.}
\end{center}
\end{figure}

Let us turn our attention to the impact of the critical current
\eqref{K_Restr} by considering the nonlinear
charge leakage term \eqref{NonLinLeakage}. We start by noting that
even for large critical currents, i.e. for $\alpha \sim 1$, the chiral
case $K\approx 0$ qualitatively  yields the same large-current
scaling behaviour which was observed previously for the linear model,
i.e. where backreaction from the current has a
significant influence on network evolution \cite{Martins:2021cid}.
This is illustrated in Fig.~\ref{Figure:NonLinA_expN}.

For critical currents $K_\text{cr}$ well below the string energy
density (with $\alpha \gg 1$), we can expect that there exist
initial conditions during the radiation era for which it should
be possible for the strings to acquire currents approaching the
critical value. Their behavior would however closely mimic that
of relativistic Nambu-Goto strings since the relevant differences
should be of order $\alpha^{-1}$. In all the cases, as in the linear
equation of state case, these currents will quickly be diluted in the
matter-dominated era and become negligible. This is again shown in
Fig.~\ref{Figure:NonLinA_expN}.

Fig.~\ref{Figure:NonLinA_expN} makes clear that the Witten model
\eqref{RealModel} essentially coincides with the much simpler linear
model for small chiralities whenever $g_{\pm}=0$ and $A_{_-}=0$.
This is by no means a trivial result, as it well known that such
an approximation is not adequate when it comes to describing 
microscopic stability issues~\cite{Peter:1992dw,Carter:1994hn}:
despite the apparent differences between the linear regime and 
the Witten model at small chiralities, they both produce the same
thermodynamical behaviour. For this reason, the key new ingredient
from the Witten model -- which is the critical current -- can be
equally introduced in the linear model to the same effect. This
still leaves open the question about the detailed nonlinear
implementation of charge leakage near the critical current
$Y\sim K_{\rm cr}$ which we have only approximately treated in
\eqref{NonLinLeakage}. However, we point out
that our results are not qualitatively changed if a different charge
leakage function is chosen, for example,
\begin{equation}
    \label{NonLinLeakageInvers}
    A_{\pm} =  
    \frac{A_{\text{const}}}{|Y- Y_{\rm cr}|} \,,
\end{equation}
where we assume the same critical value $Y_{\rm cr} = 2 / (3 \alpha)$
as before. The network current behaviour for this inverse form is 
indistinguishable from that shown in Fig.\ref{Figure:NonLinA_expN}: the
exact functional for of $A(Y)$ appears to be mostly irrelevant
provided there exists a critical current above which the leakage
becomes overwhelming ($A\to\infty$). The detailed behaviour of
this charge leakage function should be inferred from numerical
simulations, but phenomenologically the correction factor has a
negligible influence at small currents (where we can use the simple
linear model), though there are more detailed differences as we
approach the critical current.  The form of the charge leakage term,
representing a complex nonlinear process, will depend, in principle,
on the parameters of a specific superconducting cosmic string model,
see e.g.~\cite{Ibe2021, AbeHamadaSajiYoshioka}.

\subsubsection{The biased case}
\label{Bias}

Another non-trivial parameter that keeps the system of equations
\eqref{CVOS} chiral, $K_\text{sc} =0 $, is $g_{_+}$. A non vanishing
$g_{_+} \neq 0$ can arise from various underlying mechanisms through
which loops naturally form with more or less charge than the long
strings. One possible such mechanism stems from the fact that the current
momentum along the strings may smooth the
loops~\cite{CarterPeterGangui}, which would correspond to $g_{_+}<0$.
On the other hand, colliding strings create a bubble of
electromagnetic radiation that can carry away some amount of
charge~\cite{LagunaMatzner}, and this would lead to $g_{_+}>0$. Again,
numerical simulations of specific models will be required to decide
which mechanism prevails. Naturally, it is even possible that they
cancel one another, leading to the $|g|\ll 1$ case discussed above.

The result of the string network evolution with $g_{_+} \neq 0$ is
similar to cases shown in Fig.~\ref{Figure:NonLinA_expN}. It happens
because for a particular scaling regime (in our case, the
radiation-dominated epoch), there is a degeneracy between $g_{_+}$ and
$A_{_+}$ parameters: we can always tune parameters $g_{_+}$ and
$A_{_+}$ so that the scaling values of a string network evolution are
unchanged. Hence, the non-trivial value of theparameter $g_{_+}$ does not
introduce anything new since the current is non-trivial only in a
radiation-dominated epoch.

Up to this point, we did not find any difference between the full
equation of state and its linear approximation. This is due to the
fact that for the range of parameters we considered, no non-trivial
integrated chirality ensues. We now consider such a more general
situation for which the bias parameters $g_{_-}$
and $A_{_-}$ are non-negligible.

\begin{figure}[h!]
\begin{center}
\includegraphics[scale=0.38]{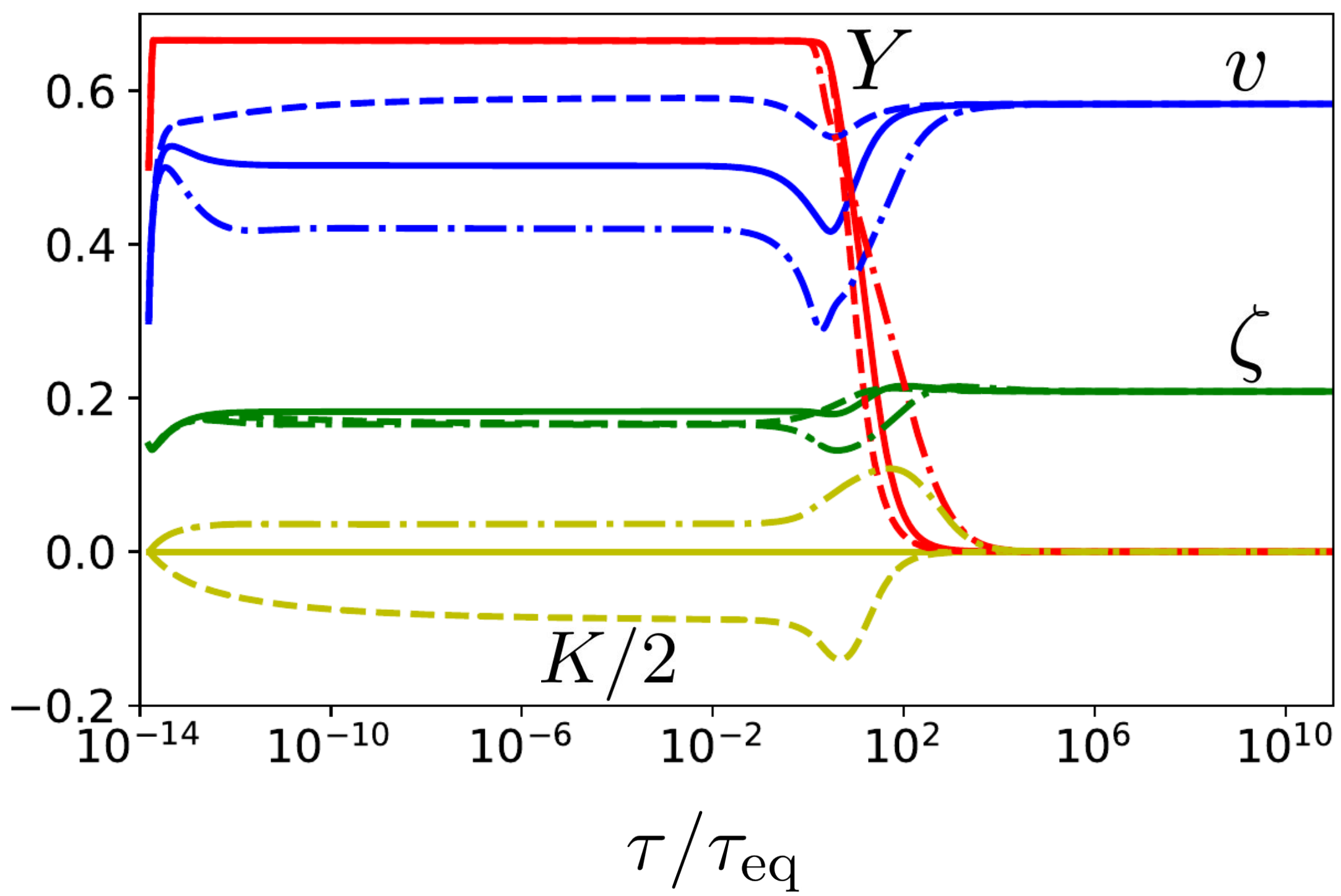}
\caption{\label{Figure:BiasG} Time evolution of the velocity $v$,
  charge $Y$, chirality $K$ and characteristic length ratio $\zeta$
  for the Witten model with $\alpha=1$, $c=0.23$, $A_{_-}=0$,
  $g_{_+}=0$ and $A_{_+}$ is given by \eqref{NonLinLeakage}, when
  $A_{\text{const}}=10^{-6}$. Parameter $g_{_-}=0$ corresponds to solid
  lines, $g_{_-}=-0.2$ for dash-dotted lines and $g_{_-}=0.2$ for
  dashed lines.}
\end{center}
\end{figure}

We just saw that the parameters $A_{_+}$ and $g_{_+}$ lead to some
amount of degeneracy in the scaling solution. A similar
situtation occurs beyween $A_{_-}$ and $g_{_-}$: one can always
change $A_{_-}$ and $g_{_-}$ to keep scaling variables the same for a particular
expansion rate. In Fig.~\ref{Figure:BiasG}, we demonstrate the
evolution of string networks for different $g_{_-}$ values. We can see
that deviation of Q leads to a non-trivial chirality in the
radiation-dominated epoch.
  
\begin{figure}
\begin{center}
\includegraphics[scale=0.38]{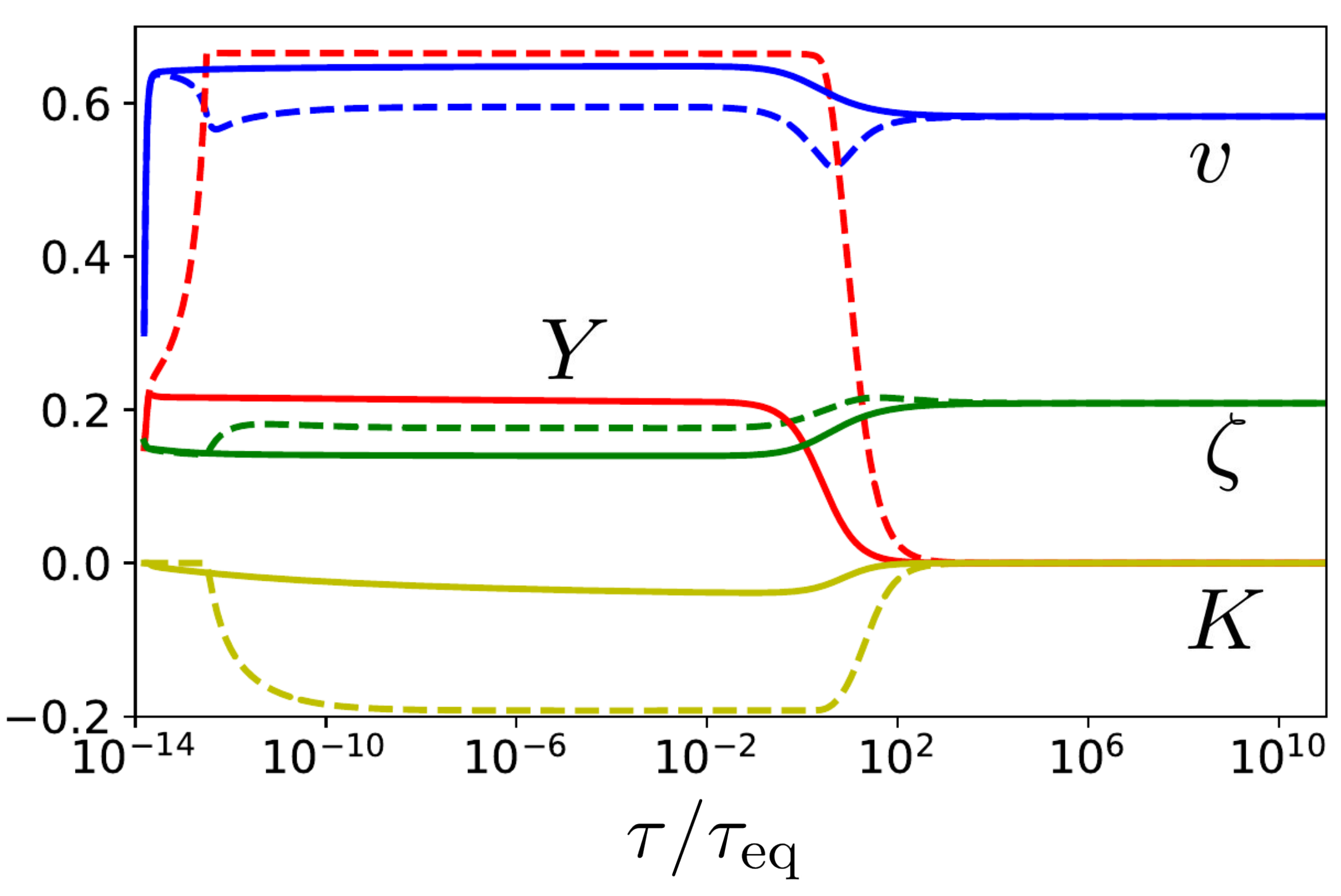}
\includegraphics[scale=0.38]{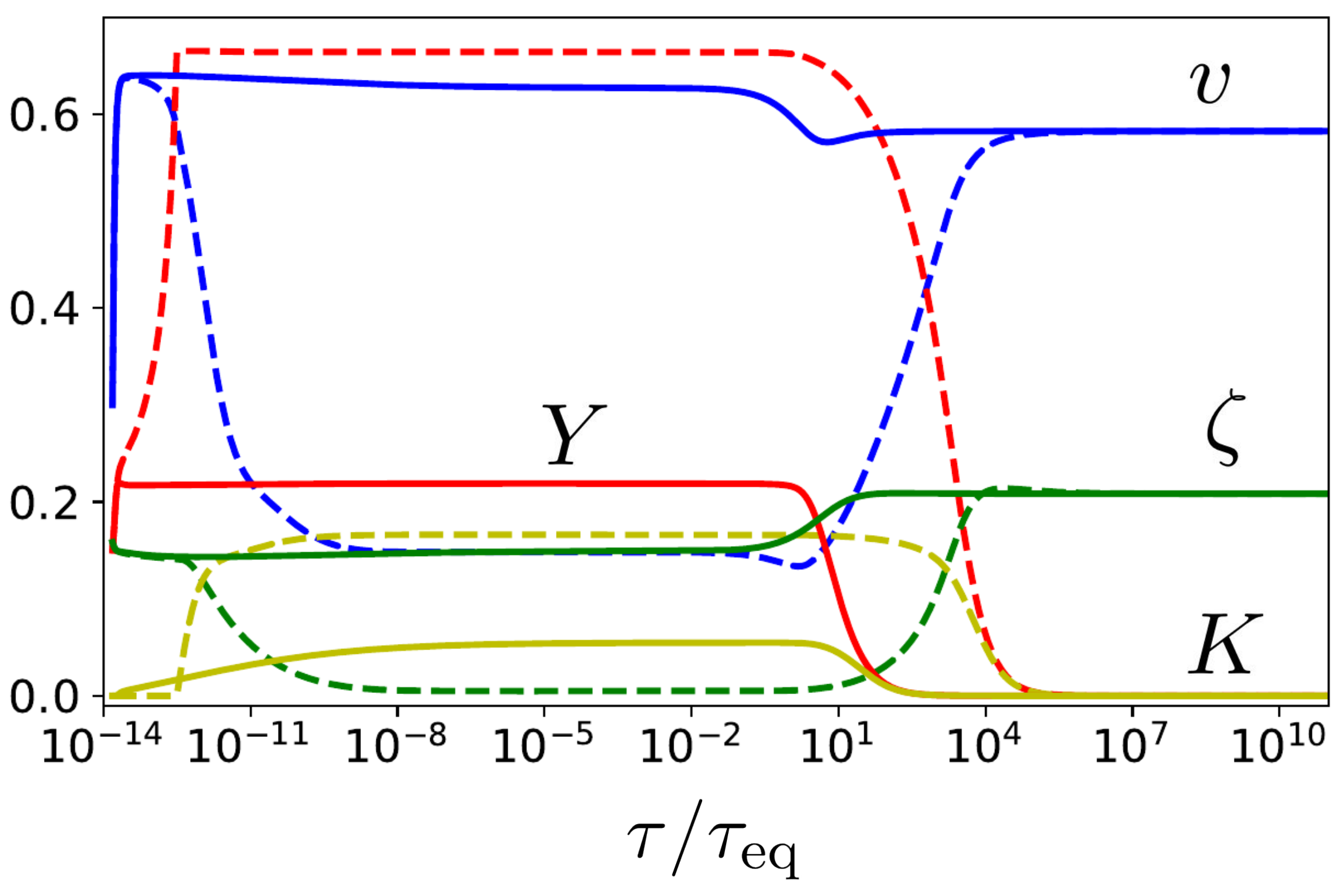}
\caption{\label{Figure:BiasN} Time evolution of the velocity $v$,
  charge $Y$, chirality $K$ and $\zeta$ through the radiation and
  matter epochs, using parameter values $c=0.23$, $g_{\pm}=0$, $A_{_-}
  = 0.1 A_{_+}$ for upper panel ($A_{_-} = -0.1 A_{_+}$ for lower
  panel), while $A_{_+}$ is given by the function
  (\ref{NonLinLeakage}) with $A_{\text{const}} = 10^{-6}$. The
  evolution is driven by the Witten equation of state
  \eqref{RealModel} with $\alpha=1$ (dashed lines) and $\alpha=3$
  (solid lines).}
\end{center}
\end{figure}

Varying the parameter $A_{_-}$, we obtain similar behaviour, i.e. the
string acquires non-trivial chirality, which we demonstrate in
Fig.~\ref{Figure:BiasN}. Also, we show that an increase of $\alpha$
leads to a decrease of the chirality variable, bringing dynamical
variables of the model with the complete equation of state closer to
the linear approximation.

\section{Conclusions}

We performed an exhaustive numerical exploration of the solutions of
the CVOS model~\cite{Martins:2020jbq} describing macroscopic
quantities (RMS velocity, correlation length, current amplitude and
charge) as functions of time for a current-carrying cosmic string
network, adopting as the underlying microscopic description the equation
of state derived~\cite{Peter:1992dw,Peter:1992ta,Carter:1994hn} in the
framework of the so-called Witten superconducting cosmic string field
theory~\cite{Witten:1984eb}. It has been previously shown that, at the
microscopic level, approximating the equation of state by its linear
expansion, even for vanishingly small currents, is not a valid
procedure as perturbation velocities do not satisfy the correct
relation~\cite{Peter:1992dw}, which in turn may yield significant
differences for vorton~\cite{Davis:1988ij}
stability~\cite{Martin:1994jp} and their cosmological
consequences~\cite{Brandenberger:1996zp}. Having studied in detail the
linear equation of state in a previous work~\cite{Martins:2021cid}, it
was therefore of interest to understand if the microscopic situation
extended to the macroscopic one, also including, for the first time, a
simple phenomenological modelling of a critical current.

The general model includes several phenomenological parameters that,
in principle, should be calibrated from high-resolution field theory
network simulations yet to be undertaken, though efforts towards this
goal are ongoing in our team. Two such parameters happen to be crucial
for the CVOS model, namely the charge leakage $A_{_+}$, representing
all possible mechanisms through which the long string network under
consideration can lose charge, and the skew $g_{_+}$ between the
charge contained on long strings and on loops. One can argue that
because there exist different mechanisms potentially increasing or
decreasing $g_{_+}$, its value may be rather small (resulting from 
the cancellation of competing effects). Similarly, at small currents 
we do not know in detail how charge may be dissipated from the
network, so the parameter $A_{_+}$ in this regime is also largely unknown. For this reason, nothing is preventing either $g_{_+}$ or $A_{_+}$ from having
a negligible effect on network evolution, while currents and charges remain well below critical.

We also studied the bias parameter $A_{_-}$, which can be non-trivial
if the leakage mechanism is not symmetric for time-like and space-like
current components. Similarly, the current carried away due to loop
production $g_{_-}$ might have a preferred channel that distorts the
symmetry between time-like and space-like parts. In both cases, the
string network deviates from chirality and the evolution becomes
distinct from the
linear approximation, the deviation, estimated of order $\alpha^{-1}$,
becoming vanishing small
for a large enough $\alpha$ value.

We have studied all possible cases for relevant parameters and found
that, for negligible charge leakage ($A_{\pm} \ll 1$) and close to
chiral initial conditions ($K_\text{ini}\approx0$), one always
recovers the linear solution, whatever the value of the bias. The same
result holds for non-negligible charge losses and skew parameter
$g_{_+}$, but vanishingly small biases
($A_{_-},g_{_-}\approx0$). Finally, when all parameters have
non-negligible values, we found that the general trends are the same
as those found in the linear model, i.e., the network approaches one of two possible scaling solutions, charged or Nambu-Goto with different scaling values, with the choice dependent on the network's initial
conditions.  When the mass of the
current-carrier becomes much smaller than the string-forming Higgs
field ($\alpha\gg 1$), however, one again recovers the linear situation (provided the current remains subcritical), so it appears to be a valid approximation for these cosmological considerations.

Scenarios with a large current (frozen) network in
the radiation era may have observational implications that are very
different from those of simpler Nambu-Goto networks. However, how easy it is to realise such scenarios remains somewhat unclear,
since the evolution of charges and currents on the network depends on their
equation of state and initial conditions (and also, in the case of
CVOS modelling, on the corresponding model parameters). In the
specific case studied in the present work, for large $\alpha \gg 1$ the small relative critical current will set a low upper limit on $Y$, thus preventing much
backreaction from the current on the network evolution.  Unless
the microscopic parameters are tuned such that the critical current has a comparable energy to the underlying string, then the general
expectation is that the network behaviour will be close to that of Nambu-Goto. In this context, the
most interesting case is when $\alpha \sim 1$, since then $Y$ can
grow much larger and there can be significant backreaction; this is also the regime suitable for vorton
formation~\cite{Battye:2021sji,Battye:2021kbd}.  A more detailed
exploration of this limit is left for subsequent work.

In a broader context, our analysis shows that the evolution of
superconducting string networks with equations of state beyond the
linear case (specifically, in the present analysis, the Witten model)
can, at least for most physically realistic cases reduce, on a
macroscopic scale, to the model with a linear equation of state.  The
conclusion is that the CVOS model offers a useful quantitative tool with which 
to start a realistic exploration of the observational constraints on
superconducting cosmic string networks.  We leave this task for a
forthcoming analysis.

\acknowledgments This work was financed by Portuguese funds through
FCT - Funda\c c\~ao para a Ci\^encia e a Tecnologia in the framework
of the projects 2022.04048.PTDC (Phi in the Sky) and 2022.03495.PTDC
(Uncovering the nature of cosmic strings).  CJM also acknowledges FCT
and POCH/FSE (EC) support through Investigador FCT Contract
2021.01214.CEECIND/CP1658/CT0001.

\bibliography{references}
\end{document}